\documentclass[journal=jacsat,manuscript=article]{achemso}
\usepackage[version=3]{mhchem}

\author{Gabriel L. Nogueira}
\affiliation{Department of Physics and Meteorology, School of Sciences, Sao Paulo State University (UNESP), Bauru, Sao Paulo, Brazil }
\email{leonardo.nogueira@unesp.br}
\author{Victor Lopez-Richard}
\affiliation{Department of Physics, Federal University of Sao Carlos (UFSCar), Sao Carlos, Sao Paulo, Brazil}
\author{Luiz A. Meneghetti Jr.}
\affiliation{Department of Physics, Federal University of Sao Carlos (UFSCar), Sao Carlos, Sao Paulo, Brazil}
\author{Fabian Hartmann}
\affiliation{Julius-Maximilians-Universität Würzburg, Physikalisches Institut and Wüzburg-Dresden Cluster of Excellence ct.qmat, Lehrstuhl für Technische Physik, Am Hubland, 97074 Würzburg, Deutschland, Würzburg, Deutchland}
\author{Carlos F. O. Graeff}
\affiliation{Department of Physics and Meteorology, School of Sciences, Sao Paulo State University (UNESP), Bauru, Sao Paulo, Brazil }
\email{carlos.graeff@unesp.br}

\title  {A New Approach to Characterize Charge Transport and Hysteresis in Perovskite Solar Cells}


\begin{document}

\begin{abstract}
Perovskite solar cells (PSCs) have emerged as a promising photovoltaic technology, already achieving efficiencies surpassing 25\%. However, effects such as hysteresis are commonly observed due to the interplay of ionic and electronic transport occurring over different timescales. Despite the widespread use of impedance spectroscopy (IS), physical interpretation in PSCs is specially challenging due to memory effects. In this study, we focus on integrating experimental data with an analytical device transport model. The PSCs under investigation were fabricated using a Cs$_{0.17}$FA$_{0.83}$Pb(I$_{0.83}$Br$_{0.17}$)$_3$ active layer between Nb$_2$O$_5$/TiO$_2$ (compact/mesoporous) and Spiro-OMeTAD. Our fully analytical charge transport model incorporates independent charge transport channels, enabling the correlation of experimental observations in both dark and under illumination. We employed IS together with various voltammetry techniques to reveal the dynamics of the transport processes, including voltage pulses and both small and large amplitude sinusoidal excitations. Although small perturbations are commonly used in IS, our findings demonstrate that large sinusoidal excitation provides new valuable insights in the transition from capacitive to inductive-like responses. The proposed model based purely on charge trapping and generation effectively captures device behavior under both small and large voltages without inductive elements, validated through accurate simulations of hysteresis and other electrical phenomena. 
\end{abstract}

\section{Introduction}
Metal halide perovskites have garnered significant attention for their potential in solar cell applications, driven by remarkable performance improvements in recent years. Perovskite solar cells (PSCs) have achieved power conversion efficiencies (PCEs) exceeding 25\%, a milestone that, in combination with their ease of fabrication, has placed them at the forefront of emergent photovoltaic technologies \cite{noman2024comprehensive}. Despite this impressive performance, perovskite-based photovoltaic devices face challenges related to stability and metastability, which complicate, for example, the accurate assessment of solar-to-electrical energy conversion efficiency \cite{koutsourakis2023investigating}. A prominent issue is the hysteresis in the current-voltage curves, which manifests as discrepancies between the forward and reverse voltage sweeps \cite{ghahremanirad2023beyond}. These effects have prompted extensive research into the fundamental properties of halide perovskites, intending to unlock their full potential for sustainable energy conversion.

These materials exhibit dual ionic-electronic transport. Typically, electronic responses occur within hundreds of nanoseconds, while the redistribution of mobile ions unfolds over seconds to minutes. Anta et al. \cite{anta2024dual} highlighted this interplay, showing how external stimuli influence ionic movement and how slow reorganization of ions can influence electronic transport. Tress et al., linked hysteresis in CH$_3$NH$_3$PbI$_3$ to ion diffusion and charge carrier collection efficiency \cite{tress2015understanding}. These dynamics are regarded as a key factor underlying hysteresis in current-voltage (J-V) measurements \cite{bisquert2024hysteresis}. The J-V characteristics of PSCs are sensitive to parameters such as voltage sweep rate, preconditioning protocols, and sweep direction. Techniques such as impedance spectroscopy (IS) have been instrumental in elucidating the interactions between ionic and electronic processes \cite{guerrero2021impedance, bisquert2021theory}. Additionally, transport models, including atomistic simulations and drift-diffusion models, have furthered our understanding of these complex phenomena \cite{anta2024dual}. 

Recent studies draw parallels between PSCs and memristors, suggesting that the resistive switching behavior commonly observed in memristors may naturally occur in PSCs due to ionic-controlled interface recombination \cite{liu2024recent} and other transport mechanisms. This underscores the suitability of using memristor models in systems involving materials with mobile ions \cite{bisquert2024hysteresis, Bisquert2023a, Bisquert2023b}. A memristor is a non-linear electrical component whose resistance depends on the history of applied voltage and current, effectively “remembering” past electrical states \cite{Chua1971, LopezRichard2022}. In PSCs, the slow redistribution of ions under external stimuli leads to resistive switching effects, analogous to memristive devices \cite{LopezRichard2022, ascoli2016history,bisquert2024inductive}. The architecture of PSC device has a strong effect on hysteresis. In planar PSCs, hysteresis is often attributed to inefficient charge transfer at the perovskite-transport layer interface, leading to charge trapping and recombination \cite{de2016ion}. Mesoporous PSCs, while exhibiting improved charge extraction compared to planar architectures, often show more pronounced hysteresis than typically observed in p-i-n devices \cite{shao2016pore}. 

In this study, we integrate multiple experiments with a transport model to investigate hysteresis in PSCs. Beyond standard electrical characterization, we employ both small AC perturbations and large amplitude voltage stimuli to probe the dynamics of transport processes. Our model provides a comprehensive framework for describing the anomalous effects observed in PSCs, moving beyond traditional equivalent circuit models.

\section{Results and discussion}

\subsection{Experimental results}

\begin{figure}
  \centering 
  \includegraphics[width=16cm]{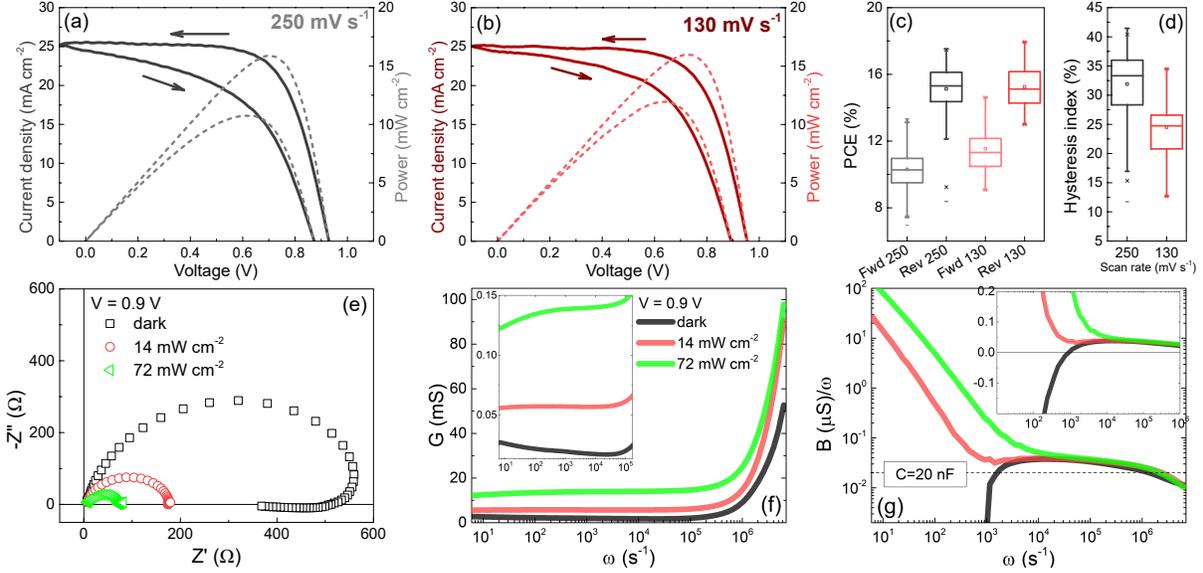}
  \caption{Representative J-V curves of the PSCs under (a) $250 mV s^{-1}$ and (b) $130 mV s^{-1}$. Box chart showing the dependance of (c) PCE and (d) hysteresis index with the scan rate. (e) Nyquist plot of the perovskite solar cell with 0.9 V bias, in the dark and under illumination. IS data converted to (f) conductance and (g) susceptance plots.}
  \label{fig1}%
\end{figure}

The J-V curves were obtained by sweeping the applied voltage over a specific range. Fig. \ref{fig1}a and Fig. \ref{fig1}b show forward and reverse bias voltage scans from representative devices, in which hysteresis is clearly observed. The average device performance yielded a PCE of 15.1\%, with $J_{\text{sc}} = 22.9\,mA/cm^2$, $V_{\text{oc}} = 0.97\,V$, and a fill factor (FF) of 67.5\%, measured at a scan rate of $250 mV s^{-1}$ in reverse direction - values consistent with prior reports \cite{lemos2023electron, fernandes2022role}. The observed counterclockwise hysteresis is classified as normal \cite{clarke2023inverted, wu2018inverted} and becomes more pronounced at faster sweep rates. In forward sweep direction, the effect of the scan rate on PCE is especially critical, as provided in Figs. \ref{fig1}c and \ref{fig1}d. 

Impedance spectroscopy is commonly used in PSCs, often complemented by equivalent circuit models. These models represent the dominant components of the system as resistors, capacitors and inductors, along with their interconnection. Fig. \ref{fig1}e illustrates the IS measurements performed at 0.9 V, close to the open-circuit condition, both in the dark and under illumination at two different light intensities (14 and 72 mWcm$^{-2}$). The dark measurement reveals a low frequency loop, consistent with previous reports of a pseudo-inductive response \cite{gonzales2022transition}. 

IS data can also be analyzed using Bode plots, where the focus is on frequency response. In Figs. \ref{fig1}f and \ref{fig1}g, we show the spectral analysis of conductance and susceptance derived from the IS data. The conductance increases proportionally with illumination intensity, while in the dark it increases as the frequency decreases. To better observe the high-frequency behavior, the susceptance is plotted as $B/\omega$, revealing its asymptotic trend toward the geometric capacitance \cite{LopezRichard2024}. Notably, the susceptance in the dark is negative at low frequencies, or pseudo inductive, consistent with the loop in the Nyquist plot \cite{bisquert2022chemical}. 

These dynamics may be associated with recombination pathways and/or electric field compensation effects \cite{hernandez2024time}. To explore these mechanisms further, we measured the current-voltage characteristics under different illumination intensities (Fig. S1a) and conducted capacitance-voltage analysis (Fig. S1c). Evaluating the open-circuit voltage, $V_{\text{oc}}$, as a function of illumination intensity provides valuable insights into recombination kinetics in PSCs. An ideality factor of $n=1$ suggests that second order (bimolecular) charge-carrier recombination dominates, whereas $n=2$ corresponds to first order (monomolecular) trap-assisted non-radiative recombination. Our devices have an ideality factor of 1.65, aligning with our previous findings using a hysteresis-free device with an inverted architecture \cite{assunccao2024interface}. These results for mesoporous PSC suggest that trap-assisted monomolecular recombination dominates near the interfaces, since hysteresis can arise from the combined effect of mobile ions and recombination near the perovskite-contact interfaces \cite{calado2016evidence}.

We performed a Mott-Schottky analysis to determine the built-in potential ($V_{bi}$) from the capacitance-voltage data (Fig. S1d). The device under investigation exhibited a $V_{bi}$ of 0.85 V, lower than that of a previously reported device with a different architecture \cite{assunccao2024interface}. A reduced built-in potential typically results in a lower open-circuit voltage, particularly when the contacts are not perfectly selective. This reduced $V_{bi}$ may explain why devices that exhibit hysteresis often display lower Voc. The $V_{bi}$ and external bias drive charge carriers toward the electrodes. However, an imbalance in the electron-hole population can cause interfacial charge accumulation, screening the electric field, and reducing charge extraction efficiency \cite{liu2019fundamental}. These findings highlight the critical role of built-in potential for ion migration and its impact on device performance \cite{liu2021correlations}. Additional details about the ideality factor and the Mott-Schottky analysis are provided in the Supplementary Information (SI).

While the slow ionic movement within the perovskite layer is widely recognized as a primary cause of the dynamic responses observed in PSCs \cite{tress2017metal}, IS and equivalent circuit models alone are insufficient to identify the ionic species involved. Our results indicated that bulk ion migration cannot fully account for hysteresis; instead, interfacial processes, which involve the capture and release of localized charges, play a significant role \cite{weber2018formation}. After diffusion, accumulated ions modify interfacial barriers and impact charge extraction at contacts \cite{sandberg2020question}. 

\subsection{Theoretical insights}

The conventional approach to analyzing IS data involves fitting the Nyquist plot with an equivalent circuit model and interpreting the physical significance of each circuit element. Yet, it has been reported that equivalent circuit representations may not adequately capture the dynamic behavior of nonlinear systems exhibiting inherent memory effects \cite{lopez2024beyond}.

In our previous work \cite{LopezRichard2024}, we proposed that a multimode framework might be better suited for a comprehensive spectral analysis as it captures the system’s complexity more effectively. However, in this study we focus on the fundamental mode ($m=1$), commonly used in traditional IS analysis, particularly for the $j_D$ term in Eq.~\ref{jt}. This approach allows for a straightforward comparison between the models.

To accurately represent the complexity of the system, we accounted for multiple concurrent drift channels, each with distinct characteristics, such as varying relaxation times and non-equilibrium carrier transfer dynamics. Under dark conditions, the experimental data was modelled using three distinct memristive channels.

\begin{figure}
  \centering 
  \includegraphics[width=12cm]{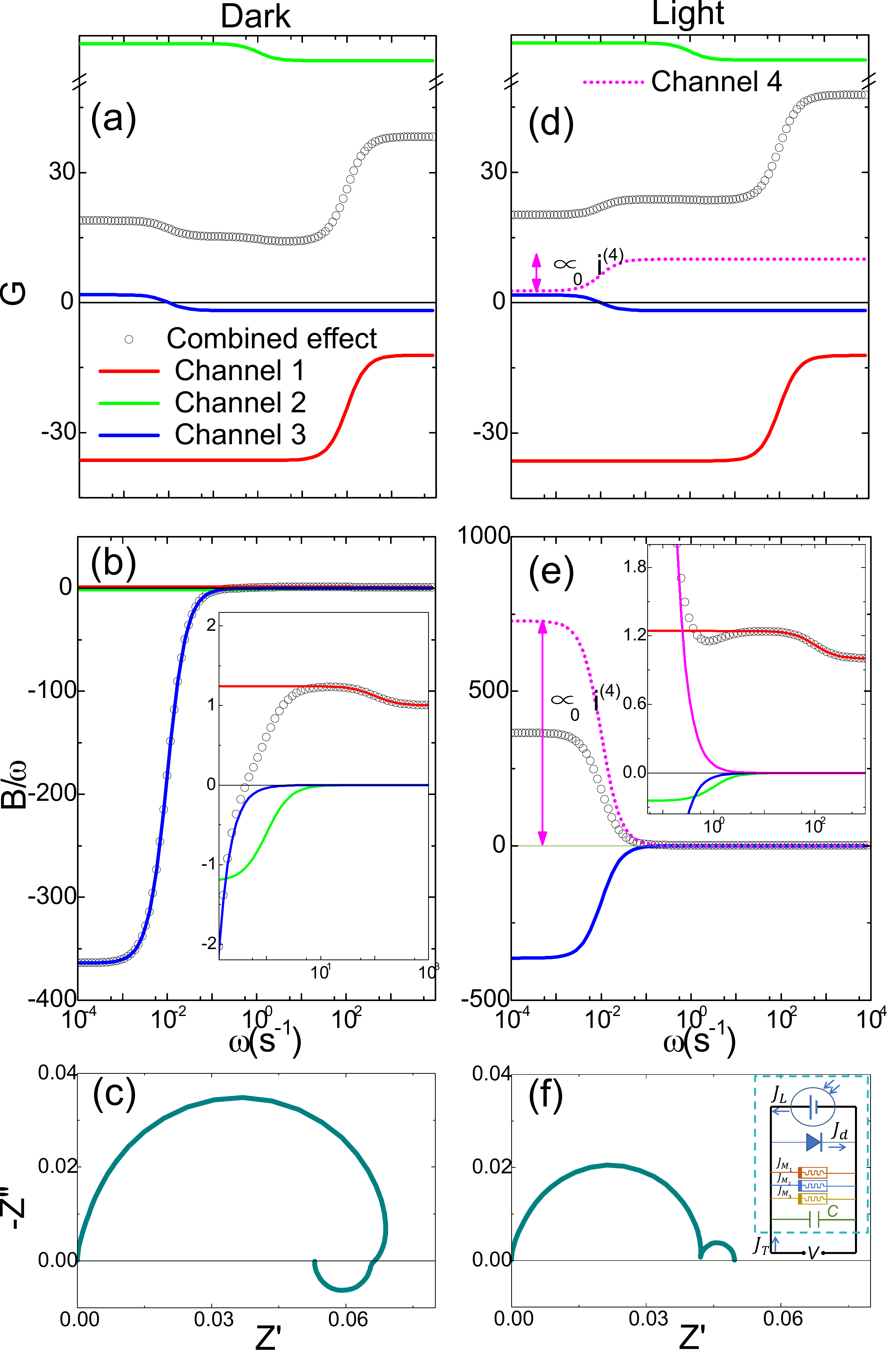}
  \caption{Simulated spectral in the dark and under illumination at a fixed DC voltage. (a,d) conductance spectrum, (b,e) susceptance, and (c,f) the corresponding Nyquist maps. The inset of (f) presents the circuit from Eq.~\ref{jt}.}
  \label{fig2}%
\end{figure}

The results from the simulated IS, incorporating contributions from these memristive components, are illustrated in Fig. \ref{fig2}. Fig. \ref{fig2}a and Fig. \ref{fig2}b display the Bode plots for conductance and susceptance, respectively, under dark conditions at constant voltage. These plots reveal the individual and combined responses of each memristive channel. Fig. \ref{fig2}c represents the corresponding Nyquist plot for the combined channel conditions under a fixed DC bias in the dark.

At low frequencies, the experimental trends - namely, the rise in conductance and the negative susceptance observed in Fig. \ref{fig1}f and Fig. \ref{fig1}g - were successfully reproduced by tuning the parameters in Eq.~\ref{gen}, specifically setting $\eta < 0$ and $\alpha = 0.9$. As demonstrated in our earlier work \cite{LopezRichard2024}, the negative susceptance originates from the reactive contribution of the system. The slight asymmetry introduced by $\alpha = 0.9$ causes the sign of the generation function to depend on the polarity of the applied bias. Expanded expressions for these terms are detailed in Ref. \cite{LopezRichard2024}.

Our model naturally captures the observed transition from capacitive to inductive response, as illustrated in Fig. \ref{fig2}. This eliminates the need to fit IS data using an equivalent circuit that includes an explicit inductive element, a practice commonly found in the literature \cite{von2022impedance}. By modeling the system with memristive components, we provide a more physically meaningful interpretation of the IS data without invoking inductive elements, which often lack physical justification in the context of solar cells.

This model effectively captures the influence of illumination on the impedance response of the solar cell, particularly its impact on the reactive components \cite{LopezRichard2024}. Illumination modifies the dynamic response by altering the effective barrier heights of the memristive channels. This effect is illustrated in Figs. \ref{fig2}d-f, where the Bode and Nyquist plots depict the impedance response with the addition of a fourth channel to simulate the experimental data under illumination, as represented by the circuit diagram shown in Fig. \ref{fig2}f and governed by Eq.~\ref{jt}.

Under illumination, the additional channel dominates the low-frequency region and is associated with photogenerated charges carriers. This results in notable effects: a decrease in conductance, a shift to positive susceptance, and the appearance of a second semicircle in the Nyquist plot. These features reflect altered charge dynamics introduced by the photogenerated carriers.

While the present model is not strictly quantitative, it successfully predicts the qualitative trends in the electrical response of perovskite solar cells. Furthermore, this framework offers flexibility for analyzing the system response under other experimental conditions, such as the electrical response to pulsed inputs or purely AC voltage excitations. By extending beyond traditional IS methods, this approach provides a versatile tool for examining additional PSCs dynamics.

\subsection{Transient response and AC stimuli}

Dynamic processes are critical for the stabilization of PSCs, particularly under large excitation \cite{bisquert2024hysteresis}. We propose an alternative protocol to IS, by utilizing a two-step methodology: (i) a voltage pulse to identify key time constants, and (ii) a sinusoidal voltage input instead of the traditional triangular voltage ramp. The sinusoidal stimulus is used for detailed frequency-dependent dynamics, analogous to but complementary to conventional IS. Despite the illumination effects discussed in previous sections, we focus here on dark conditions, where the inductive-like characteristic is observed. Additionally, in the absence of light, charge carriers are injected solely from the contacts, making the system less susceptible to photodegradation effects \cite{gonzales2022transition}.

\begin{figure}
  \centering 
  \includegraphics[width=15cm]{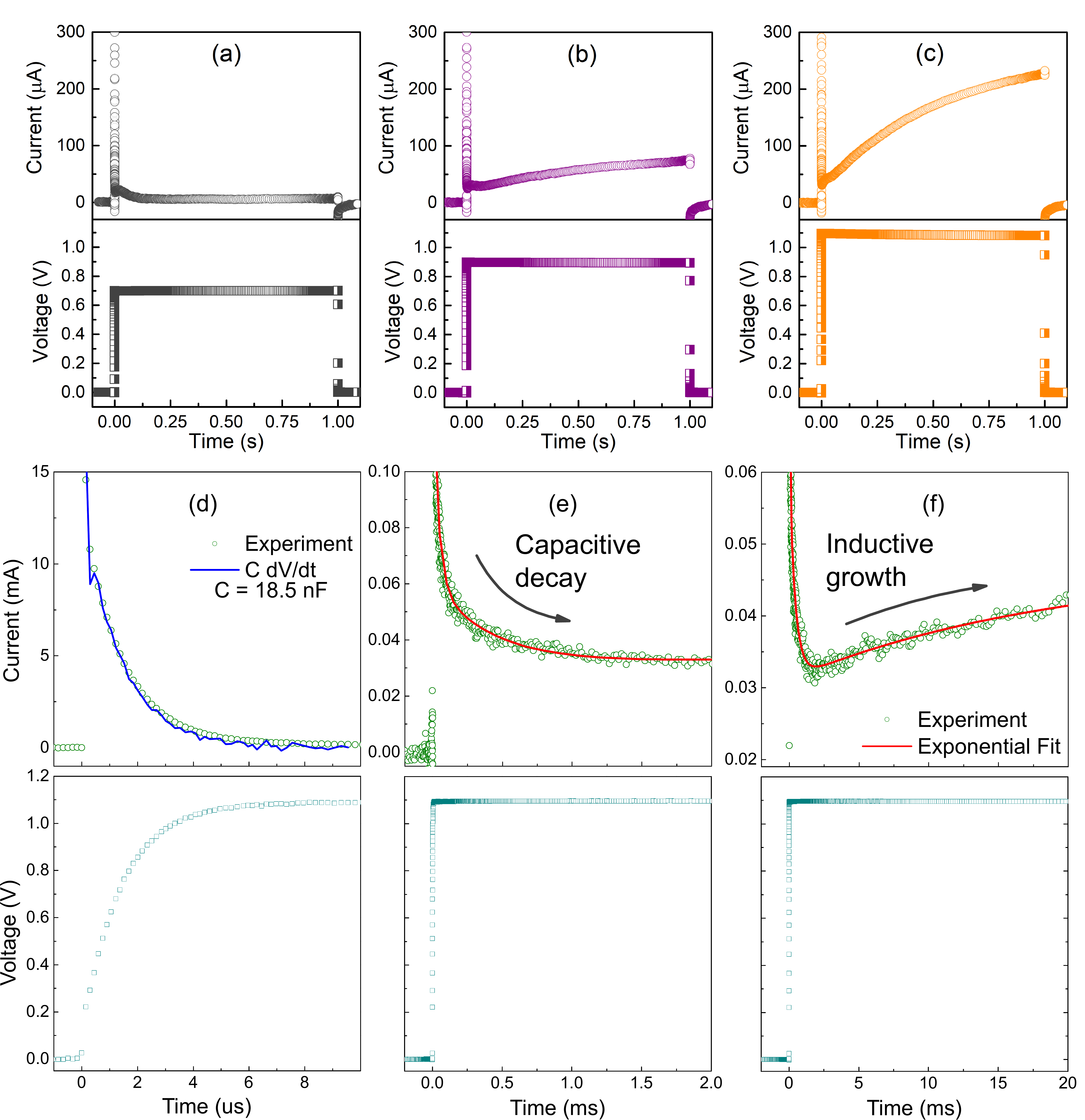}
  \caption{Transient response in the dark for a PSC after a voltage pulse of (a) 0.7 V, (b) 0.9 V and (c) 1.1 V. The transient of the 1.1 V pulse is shown for three different time scales: (d) 10 us, (e) 2 ms and (f) 20 ms. The solid red line represents the calculated $CdV/dt$ in plot (d), and an exponential fit in plots (e) and (f).}
  \label{fig3}%
\end{figure}

Fig. \ref{fig3} illustrates the current response to 1 second voltage pulses under different amplitudes in the dark. Under a 0.7 V pulse (Fig. \ref{fig3}a), the device exhibits an initial microsecond-scale spike followed by a steady-state current. When the bias exceeds the built-in potential ($V_{bi}$), the current spike is followed by a gradual increase (Fig. \ref{fig3}b). This so-called “negative transient spike” is typically both voltage- and light -dependent (Fig. \ref{fig3}c) \cite{Bisquert2023b, h2024accelerating}. 

Figures \ref{fig3}d-f decompose the response to a 1.1 V pulse into three distinct timescales, corresponding to capacitive and “inductive” responses as commonly described in the literature \cite{hernandez2022negative}. The transient response comprises an initial fast spike, a subsequent current decay, and a slower current increase that stabilizes over time. The fastest process (~1 $\mu$s), shown in Fig. \ref{fig3}d, reflects a pure capacitive effect. Fitting this response with the voltage derivative with respect to the current yields a capacitance of $\approx$18.5 nF, consistent with the value extracted via IS in the frequency range of 10000 and 100000 Hz. Figures \ref{fig3}e and \ref{fig3}f illustrate the subsequent current decay and the onset of slower current growth, respectively. 

The current behavior was modelled using a sum of three exponential terms: \begin{math}f(x)=A + b_1\exp(-\frac{t}{\tau_1}) + b_2\exp(-\frac{t}{\tau_2}) + b_3\exp(-\frac{t}{\tau_3})\end{math}. Despite being usually related to the capacitive and inductive elements, we assume that the decay and growth exponents correspond to a competition of trapping rates and carrier generation, as described by Eq.~\ref{gen} \cite{LopezRichard2024}. A capacitance plot and a table summarizing the exponential parameters are provided in the SI. We highlight the third term, where the negative value of b$_3$ indicates that carrier generation processes become predominant above the characteristic time ($\tau_3$) of about 20 ms. 

Fast dynamics, typically on timescales from $10^{-12}\, s$ to $10^{-6}\, s$, are attributed to electronic processes, while slower dynamics, lasting over 1 second, are associated with electrochemical or ionic processes \cite{von2022impedance}. The microsecond-scale capacitive response observed in our experiments (Fig. \ref{fig3}d) may be linked to recombination processes either at the interfaces or within the perovskite bulk. As discussed in Section 3.1, an ideality factor of 1.65 suggests that while trap-assisted recombination contributes to the overall response, it is not the dominant recombination mechanism. 

We infer that the slower processes, exceeding 1 s, are associated with ionic motion. This assignment is supported by comparing the timescale of the current decay in Fig. \ref{fig3}e, 0.1 to 2 ms, with the negative susceptance below 1000 Hz in Fig. \ref{fig1}g. The origin of these slow dynamics may involve relatively fast Li$^+$ ions from the spiro transport layer, slower movement of heavier ions between the perovskite and transport layers, or the formation of ion vacancies at grain boundaries in the perovskite film \cite{o2017measurement}. It is important to note that the presence of transients imposes additional challenges for spectral analysis as highlighted by Lopez-Richard et al. \cite{lopez2024beyond}.

\begin{figure}
  \centering 
  \includegraphics[width=15cm]{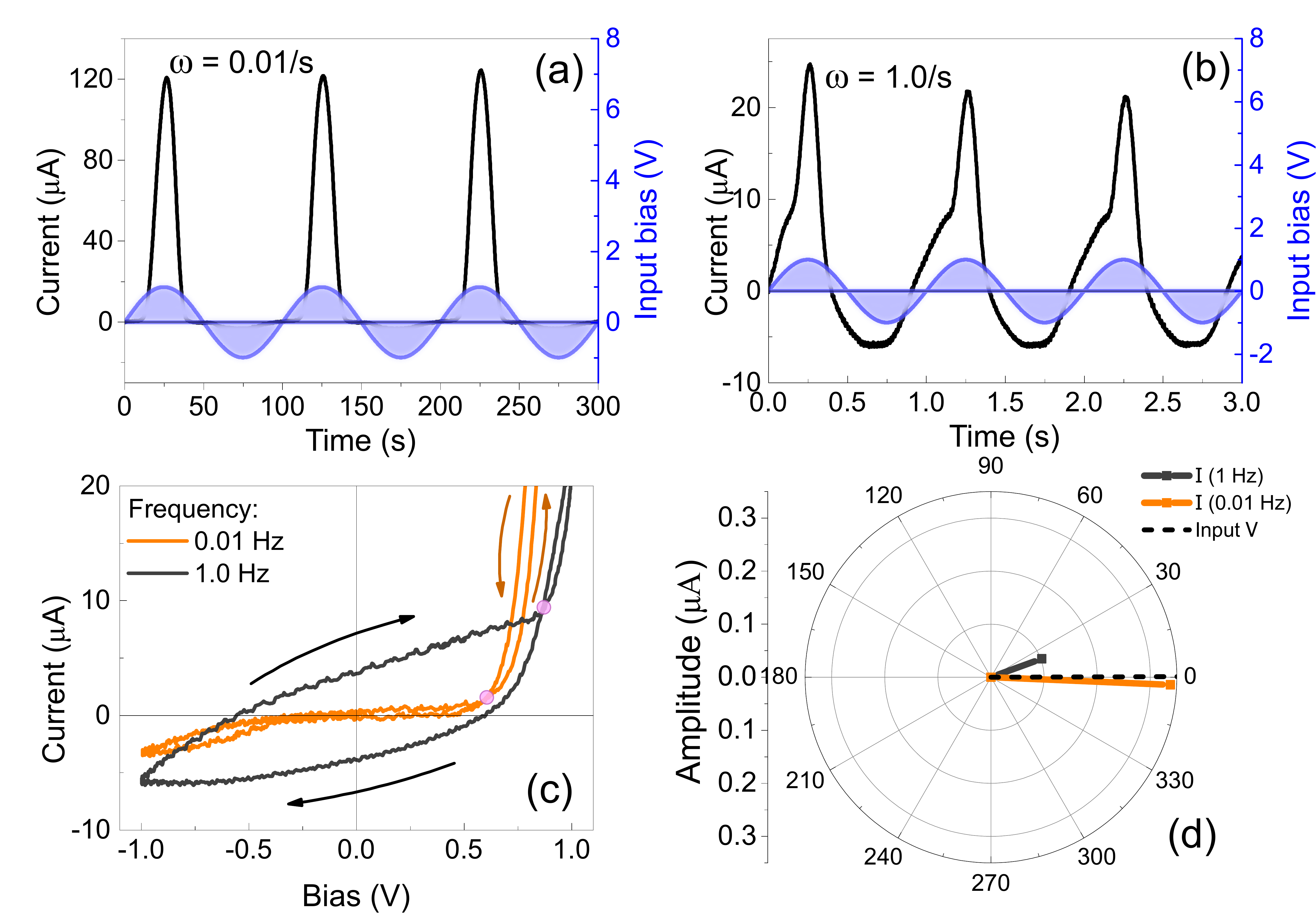}
  \caption{Experimental current and voltage time response in the dark under a sinusoidal voltage of 1V applied to the PSC with (a) 0.01 Hz and (b) 1 Hz. (c) Current-voltage characteristics for the two frequencies. The arrows indicate the scan direction, while the pink circles highlight the transition from a capacitive (clockwise) to an inductive-like response (counter-clockwise). (d) The polar map extracted from the FFT represents the capacitive and inductive-like response on the current phase compared to the input voltage.}
  \label{fig4}%
\end{figure}

From the dark transient response, our focus is on the current response at voltage biases around 1V and processes occurring on milliseconds. Figure \ref{fig4} displays the steady-state current-voltage characteristics in the dark, measured using a sinusoidal voltage input of 1 V at two distinct frequencies: 0.01 Hz and 1 Hz, Figs. \ref{fig4}a-b. The Sinus-J-V measurement analysis is detailed in Fig. \ref{fig4}c. In this figure, a frequency dependent hysteresis effect is visible. Depending on the voltammetry and IS data, hysteresis could usually be classified as capacitive or inductive. For instance, the counterclockwise hysteresis and the low-frequency loop at Nyquist plot are both indicative of an inductive-like response \cite{bisquert2024inductive}. In previous work \cite{LopezRichard2024}, we observed that the frequency tunning of clockwise and counter-clockwise current-voltage loops results from combinations of nonequilibrium carrier trapping and generation, respectively. These behaviors can be described as capacitive or inductive based on the apparent anticipation or delay of the current with respect to the voltage sweep. Therefore, the low frequency curve in Fig. \ref{fig4} has an inductive-like response attributed to nonequilibrium charges, coexisting with a capacitive transport component that is delimited by a transition point \cite{gonzales2022transition}. At higher frequencies, the influence of nonequilibrium charges weakens and the capacitive response dominates. As marked by the pink circles in Fig. \ref{fig4}c, the transition points also depend on the frequency, being at 0.61 V and 0.87 V for the low and high frequency curves, respectively. 

The relationship between frequency ($f$) and scan rate ($\nu$) can be approximated as $f=\frac{\nu}{2 \Delta E}$ \cite{schalenbach2021double}. The frequencies of 0.01 Hz and 1 Hz correspond to a triangular wave with scan rates of $40\, mV s^{-1}$ and $4\, V s^{-1}$, respectively. This trend aligns with the observations by Thiesbrummel et al. \cite{thiesbrummel2024ion}, who reported a similar frequency-dependent hysteresis effect between 0.1 and 100 V s$^{-1}$. Thus, hysteresis is minimal at slower scan rates (low frequency) and becomes more pronounced at higher scan rates. At further higher scan rates, it would be expected that the ion-freeze effect mitigates hysteresis because ions cannot respond to rapid perturbations. 

The map of polar coordinates in Fig. \ref{fig4}d shows how the phase shift between voltage and current aligns with the capacitive and inductive hysteresis response observed in Fig. \ref{fig4}c. The plot was obtained by using the Fast Fourier Transform (FFT) to convert the time-resolved sinusoidal data to the frequency domain, in which the current amplitude is related to the first harmonic of Fig. S2. The dashed line marks the input voltage phase, arbitrarily positioned at 0$^o$. The excitation frequency significantly influences the current phase. At 1 Hz, the current phase is 20$^o$, while at 0.01 Hz the phase is shifted to -2.4$^o$. Fig. S4e presents the map that highlights the range of 330$^o$ to 30$^o$. This phase analysis supports the conclusion that the dark transients exhibit inductive-like responses at slower timescales and capacitive responses at faster timescales, as demonstrated in Fig. \ref{fig3}. 

\begin{figure}
  \centering 
  \includegraphics[width=15cm]{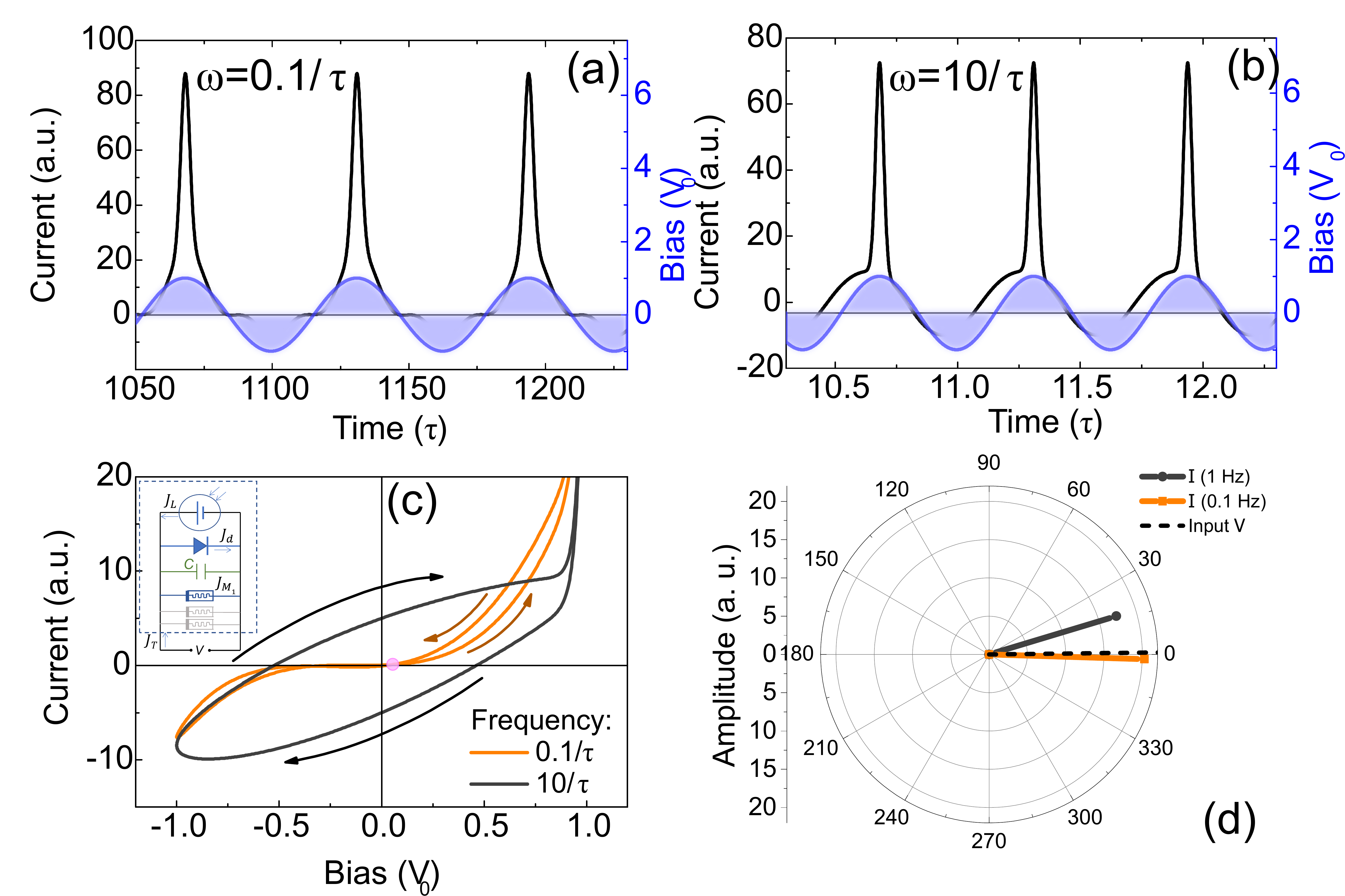}
  \caption{Simulated current and voltage time response in the dark under an applied sinusoidal voltage of 1 V with (a) 0.1/$\tau$ and (b) 10/$\tau$. (c) Current-voltage characteristic for the two frequencies. The arrows indicate the scan direction, while the pink circle highlights the transition from a capacitive (clockwise) to an inductive-like response (counter-clockwise).  (d) The polar map extracted from the FFT represents the capacitive and inductive-like response on the current phase compared to the input voltage. Inset in (c) is the circuit diagram of Eq.~\ref{jt}. }
  \label{fig5}%
\end{figure}

Our model, considering only the fundamental harmonic and a single drift channel, is represented by the circuit diagram shown in the inset of Fig. \ref{fig5}c. The simulated curves assume a scenario where a stationary DC voltage is superimposed on an AC voltage component. Figures 5a-b show the current response for two AC voltage conditions. The Sinus-J-V analysis and the polar coordinate map in Figs. \ref{fig5}c and \ref{fig5}d closely match the experimental data, validating the direct comparison between model and experiments. At higher frequency, the current exhibits a phase of 16.8$^o$, while at lower frequency the phase shifted to -1.7$^o$. The requirement for only one drift channel suggests that the lag in output current under faster perturbations corresponds to a dominant process with a characteristic response time. The simulated curves also show the transition point for the slower sweep, highlighting the inductive-like character at such condition, alternatively visualized by the polar map of Fig. \ref{fig5}d. Therefore, our model successfully simulates the transition from capacitive to inductive-like without using equivalent circuits with inductive elements. Instead, the simple analytical approach is based purely on charge trapping and generation. Additional details, including Fourier transform outputs for experimental and simulated data, are provided in the SI.

Further work would be focused on the analysis of sinus-J-V and FFT under illumination, which will enable better comparison to realistic solar cell measurement protocols. The advantage of using sinus-J-V is that capacitive hysteresis manifests itself as an oval-shaped current-voltage curve. The large widening corresponds to the characteristic response time of the dominant charge transport mechanism or species related to trapping phenomena. On the other hand, the inductive-like response manifest by changing the loop direction above a specific transition point. In our case, the characteristic response time for the charge generation is from tens of milliseconds to seconds, supporting ion migration as the primary non-electronic cause of hysteresis. This is consistent with the mesoporous architecture and perovskite composition, since the iodide ions are relatively small compared to the lattice structure and require less activation energy for migration compared to other intrinsic ions such as Cs$^+$, FA$^+$ or Br$^-$ \cite{zuo2023ion}. Although the exact migrating species remains debated \cite{gollino2024unveiling, yuan2023targeted,tammireddy2024hysteresis}, current understanding suggests that these ions reduce charge collection efficiency by screening the internal electric field \cite{thiesbrummel2024ion}. Therefore, sinus-J-V combined with the proposed analytical model is particularly helpful as it captures the frequency-dependent dynamics of ionic migration more accurately than triangular voltage sweeps and offers deeper insights into the nonlinear process beyond the limitations of conventional small signal impedance spectroscopy method.

\section{Conclusion}
In conclusion, we present a unified theoretical framework for analyzing charge transport in PSCs, grounded in experimental data from a standard n-i-p mesoporous device. Using three independent memristive transport channels, we effectively numerically derived the IS in the dark. Under illumination, a fourth memristive channel was critical to accurately simulate the IS response.

In addition to IS, our model can be used to simulate the electrical response of PSC under both pulsed and sinusoidal large stimuli. The transient response to voltage pulses in the dark revealed at least three distinct processes: a rapid capacitive spike, an initial current decay, and a subsequent long-term increase. The first effect occurs at 10 $\mu$s, corresponding to a capacitance of around 18.5 nF. The second process dominates around 2 ms, while the third becomes significant when it exceeds 20 ms. 

The dark Sinus-J-V through FFT analysis presents the system transition from capacitive to pseudo inductive dynamics, based on amplitude and phase, yielding the trapping and generation of charges as the predominant mechanism. At a low frequency of 0.01 Hz, the input voltage and the current response presented a phase of -2.41$^o$, resulting in minimal capacitive hysteresis. In contrast, at 1.0 Hz, a phase lag of 20$^o$ between the input voltage and the current response, corresponding to an increase in hysteresis. The simulated results closely mirrored this behavior, predicting a phase lag of 17$^o$ at the higher frequency.

Notably, our model accurately predicted the electrical response of real devices without relying on conventional equivalent circuit models. This approach offers a novel and comprehensive understanding of hysteresis phenomena in PSCs and paves the way for more robust and versatile methods to characterize these systems.

\section{Experimental}

\subsection{Experimental details}

Fluorine-doped Tin Oxide (FTO) coated glass substrates ($7\, \Omega.\text{sq}^{-1}$, Sigma Aldrich) were patterned and sequentially cleaned. After drying with N$_2$, they were treated with UV-Ozone (Ossila Ltda) for 15 minutes to ensure optimal surface cleanliness. A compact Nb$_2$O$_5$ layer was deposited via reactive magnetron sputtering in a Kurt J. Lesker System. The deposition was conducted at 500 $^o$C with 240 W plasma power. The chamber was maintained at a pressure of $5.0 \times 10^{-3}$ Torr, with a 40 sccm flow of high-purity argon and 3.5 sccm of oxygen. A mesoporous TiO2 solution (150 mg/ml in ethanol) was spin-coated on top of the Nb$_2$O$_5$ compact layer. Both compact and mesoporous layers were annealed at 550 C for 60 minutes. FTO/Nb$_2$O$_5$/TiO$_2$ substrates were transferred to a nitrogen-filled glovebox for device fabrication. A Cs$_{0.17}$FA$_{0.83}$Pb(I$_{0.83}$Br$_{0.17}$)$_3$ double-cation mixed-halide perovskite solution was prepared by combining two separately prepared solutions: one iodide-based (1.2 M PbI$_2$, 1.0 M FAI, 0.2 M CsI in DMF:DMSO, 4:1 v/v) and one bromide-based (1.2 M PbBr$_2$, 1.2 M FABr in DMF:DMSO, 4:1 v/v). The perovskite film was deposited via spin-coating: first at 1000 rpm for 10 s, then at 6000 rpm for 25 s, with 200 $\mu$L of chlorobenzene (anti-solvent) dropped after 20 s. The perovskite film was first annealed at 60$^o$C for 3 minutes and then at 120$^o$C for 30 minutes to promote crystallization. The hole transporting layer (HTL) was based on Spiro-OMeTAD (73 mg/mL in chlorobenzene) doped with 29 $\mu$L of 4-tert-butyl pyridine, 29 uL of FK209 cobalt complex solution (300 mg/mL in acetonitrile), and 18 $\mu$L of lithium bis(trifluoromethanesulfonic)imide (Li-TFSI) solution (520 mg/mL in acetonitrile). Finally, an 80 nm gold electrode was thermally evaporated onto the device integrated to the Glove Box. More details on the depositions can also be found in the previous publication \cite{lemos2023electron, fernandes2022role}.

J-V curves were measured using a Keithley 2400 source measurement unit under simulated sunlight (AM 1.5G,$100\,mW\,cm^{-2}$) from a Newport Class AAA solar simulator (model 94023AU), calibrated with a certified silicon solar cell. Additional electrical characterization was performed using the PAIOS system (Fluxim AG). Techniques included staircase sweep J-V, where data points are acquired step-by-step for higher precision, and dark injection transients (DIT), which apply a voltage step and measure the resulting transient current. Sinus-IV measurements applied a sinusoidal voltage, plotting the resulting current versus voltage. Impedance spectroscopy (IS) involved applying a small AC signal (50 mV) superimposed on a DC bias and measuring current across a range of frequencies.

\subsection{Analytical model}

The analytical model employed in this study builds on our previous work \cite{LopezRichard2024}. In summary, for an arbitrary bias voltage, $V$, the total current density, $j_T$, is decomposed into three independent components,
\begin{equation}
    j_T=j_D+\frac{C_g}{A} \frac{dV}{dt}+\sum_i j_{M_i},
    \label{jt}
\end{equation}
where $j_D$ represents the diode current density. This term represents the current derived from minority carrier diffusion at the boundaries of the depletion (or intrinsic) region
\begin{equation}
j_D=e D_n \frac{\partial n}{\partial z}|_{z= -\frac{\Delta}{2}}-e D_p \frac{\partial p}{\partial z}|_{z= \frac{\Delta}{2}}, 
    \label{jd}
\end{equation}
of width $\Delta$. The axis origin is set at the mid-point of the depletion region assuming uniform electron and hole current components \cite{Green1982}. 

The second term in Eq.~\ref{jt} describes the displacement current contribution due to the geometric capacitance, $C_g$, of the device with area $A$. The third term models memory-related contributions, including ionic channels, fluctuations, and potential leakage pathways (labelled with the sub-index $i$). It is expressed as, 
\begin{equation}
j_{M_i}=\gamma_i (N_i^0+\delta N_i)V,
   \label{jm}
\end{equation}
where $\gamma_i = \frac{e \mu_i}{A\Delta^2}$ is the ionic mobility term along $\Delta$, and $N_i^0 + \delta N_i$ represents the fluctuating non-equilibrium carrier population around $N_i^0$ with a relaxation time $\tau_i$~\cite{Silva2022,Paiva2022},
\begin{equation}
    \frac{d \delta N_i}{d t}=-\frac{\delta N_i}{\tau_i}+g_i(V).
    \label{dn}
\end{equation}

According to \cite{Silva2022}, the carrier generation or trapping rate is described by a transfer function
\begin{equation}
    g(V)=\frac{i_0}{\eta}\left[e^{-\eta_L \frac{eV}{k_B T}}+e^{\eta_R \frac{eV}{k_B T}} -2 \right],
    \label{gen}
\end{equation}
where $i_0=\frac{4 \pi m^* A}{(2 \pi \hbar)^3} (k_B T)^2  e^{-\frac{E_B}{k_B T}}$, $\eta_L=\alpha \eta/(1+\alpha)$, and $\eta_R= \eta/(1+\alpha)$. Here, $\alpha \in [0,\infty)$ characterizes the symmetry of the carrier transfer with respect to the local bias voltage drop. The case $\alpha=1$ corresponds to symmetric transfer, while $\eta>0$ and $\eta<0$ represent pure generation and pure trapping scenarios, respectively. The terms $j_D$ and $j_{M_i}$ from Eq.~\ref{jt} were thoroughly expanded in Ref. \cite{Paiva2022}. 

For this study, both AC and DC conditions are addressed using the general voltage expression $V = V_S + V_0 \cos(\omega t)$, encompassing cyclic voltammetry and IS characterizations for arbitrary $V_S$. The combination of these terms consistently reproduces memory effects \cite{LopezRichard2022, Silva2022,Paiva2022}, providing a framework to correlate current-voltage characteristics with IS measurements.

\begin{acknowledgement}

This study was financed in part by the Coordenacão de Aperfeicoamento de Pessoal de Nivel Superior - Brazil (CAPES) and the Conselho Nacional de Desenvolvimento Científico e Tecnologico - Brazil (CNPq) (311536/2022-0; 408041/2022-6), and in part by Fundacão de Amparo a Pesquisa do Estado de Sao Paulo (FAPESP) (20/12356-8; 22/10998-8; 23/01117-0). The authors also thank the support of FINEP (Grant: 01.22.0289.00 (0034-21)) and UNESP.

\end{acknowledgement}


\begin{thebibliography}{50}

\bibitem[{Noman et~al.(2024)Noman, Khan and Jan}]{noman2024comprehensive}
\bibinfo{author}{Noman, M.}, \bibinfo{author}{Khan, Z.}, \bibinfo{author}{Jan, S.T.}, \bibinfo{year}{2024}.
\newblock \bibinfo{title}{A comprehensive review on the advancements and challenges in perovskite solar cell technology}.
\newblock \bibinfo{journal}{RSC advances} \bibinfo{volume}{14}, \bibinfo{pages}{5085--5131}.
%Type = Article

\bibitem[{Koutsourakis et~al.(2023)Koutsourakis, Worsley, Spence, Blakesley, Watson, Carnie and Castro}]{koutsourakis2023investigating}
\bibinfo{author}{Koutsourakis, G.}, \bibinfo{author}{Worsley, C.}, \bibinfo{author}{Spence, M.}, \bibinfo{author}{Blakesley, J.C.}, \bibinfo{author}{Watson, T.M.}, \bibinfo{author}{Carnie, M.}, \bibinfo{author}{Castro, F.A.}, \bibinfo{year}{2023}.
\newblock \bibinfo{title}{Investigating spatial macroscopic metastability of perovskite solar cells with voltage dependent photoluminescence imaging}.
\newblock \bibinfo{journal}{Journal of Physics: Energy} \bibinfo{volume}{5}, \bibinfo{pages}{025008}.
%Type = Article

\bibitem[{Anta et~al.(2024)Anta, Oskam and Pistor}]{anta2024dual}
\bibinfo{author}{Anta, J.A.}, \bibinfo{author}{Oskam, G.}, \bibinfo{author}{Pistor, P.}, \bibinfo{year}{2024}.
\newblock \bibinfo{title}{The dual nature of metal halide perovskites}.
\newblock \bibinfo{journal}{The Journal of Chemical Physics} \bibinfo{volume}{160}.
\newblock {DOI: 10.1063/5.0190890}.
%Type = Article

\bibitem[{Ghahremanirad et~al.(2023)Ghahremanirad, Almora, Suresh, Drew, Chowdhury and Uhl}]{ghahremanirad2023beyond}
\bibinfo{author}{Ghahremanirad, E.}, \bibinfo{author}{Almora, O.}, \bibinfo{author}{Suresh, S.}, \bibinfo{author}{Drew, A.A.}, \bibinfo{author}{Chowdhury, T.H.}, \bibinfo{author}{Uhl, A.R.}, \bibinfo{year}{2023}.
\newblock \bibinfo{title}{Beyond protocols: Understanding the electrical behavior of perovskite solar cells by impedance spectroscopy}.
\newblock \bibinfo{journal}{Advanced Energy Materials} \bibinfo{volume}{13}, \bibinfo{pages}{2204370}.
%Type = Article

\bibitem[{Tress et~al.(2015)Tress, Marinova, Moehl, Zakeeruddin, Nazeeruddin and Gr{\"a}tzel}]{tress2015understanding}
\bibinfo{author}{Tress, W.}, \bibinfo{author}{Marinova, N.}, \bibinfo{author}{Moehl, T.}, \bibinfo{author}{Zakeeruddin, S.M.}, \bibinfo{author}{Nazeeruddin, M.K.}, \bibinfo{author}{Gr{\"a}tzel, M.}, \bibinfo{year}{2015}.
\newblock \bibinfo{title}{Understanding the rate-dependent j--v hysteresis, slow time component, and aging in ch 3 nh 3 pbi 3 perovskite solar cells: the role of a compensated electric field}.
\newblock \bibinfo{journal}{Energy \& Environmental Science} \bibinfo{volume}{8}, \bibinfo{pages}{995--1004}.
%Type = Article

\bibitem[{Bisquert(2024a)}]{bisquert2024hysteresis}
\bibinfo{author}{Bisquert, J.}, \bibinfo{year}{2024}a.
\newblock \bibinfo{title}{Hysteresis, impedance, and transients effects in halide perovskite solar cells and memory devices analysis by neuron-style models}.
\newblock \bibinfo{journal}{Advanced Energy Materials} , \bibinfo{pages}{2400442}.
%Type = Article

\bibitem[{Guerrero et~al.(2021)Guerrero, Bisquert and Garcia-Belmonte}]{guerrero2021impedance}
\bibinfo{author}{Guerrero, A.}, \bibinfo{author}{Bisquert, J.}, \bibinfo{author}{Garcia-Belmonte, G.}, \bibinfo{year}{2021}.
\newblock \bibinfo{title}{Impedance spectroscopy of metal halide perovskite solar cells from the perspective of equivalent circuits}.
\newblock \bibinfo{journal}{Chemical Reviews} \bibinfo{volume}{121}, \bibinfo{pages}{14430--14484}.
%Type = Article

\bibitem[{Bisquert et~al.(2021)Bisquert, Guerrero and Gonzales}]{bisquert2021theory}
\bibinfo{author}{Bisquert, J.}, \bibinfo{author}{Guerrero, A.}, \bibinfo{author}{Gonzales, C.}, \bibinfo{year}{2021}.
\newblock \bibinfo{title}{Theory of hysteresis in halide perovskites by integration of the equivalent circuit}.
\newblock \bibinfo{journal}{ACS Physical Chemistry Au} \bibinfo{volume}{1}, \bibinfo{pages}{25--44}.
%Type = Article

\bibitem[{Liu et~al.(2024)Liu, Zeng, Chen and Liu}]{liu2024recent}
\bibinfo{author}{Liu, S.}, \bibinfo{author}{Zeng, J.}, \bibinfo{author}{Chen, Q.}, \bibinfo{author}{Liu, G.}, \bibinfo{year}{2024}.
\newblock \bibinfo{title}{Recent advances in halide perovskite memristors: From materials to applications}.
\newblock \bibinfo{journal}{Frontiers of Physics} \bibinfo{volume}{19}, \bibinfo{pages}{23501}.
%Type = Article

\bibitem[{Bisquert(2023)}]{Bisquert2023a}
\bibinfo{author}{Bisquert, J.}, \bibinfo{year}{2023}.
\newblock \bibinfo{title}{Current-controlled memristors: Resistive switching systems with negative capacitance and inverted hysteresis}.
\newblock \bibinfo{journal}{Phys. Rev. Appl.} \bibinfo{volume}{20}, \bibinfo{pages}{044022}.
\newblock {DOI: 10.1103/PhysRevApplied.20.044022}.
%Type = Article

\bibitem[{Bisquert et~al.(2023)Bisquert, Bou, Guerrero and Hernández-Balaguera}]{Bisquert2023b}
\bibinfo{author}{Bisquert, J.}, \bibinfo{author}{Bou, A.}, \bibinfo{author}{Guerrero, A.}, \bibinfo{author}{Hernández-Balaguera, E.}, \bibinfo{year}{2023}.
\newblock \bibinfo{title}{Resistance transient dynamics in switchable perovskite memristors}.
\newblock \bibinfo{journal}{APL Machine Learning} \bibinfo{volume}{1}, \bibinfo{pages}{036101}.
\newblock {DOI: 10.1063/5.0153289}.
%Type = Article

\bibitem[{Chua(1971)}]{Chua1971}
\bibinfo{author}{Chua, L.}, \bibinfo{year}{1971}.
\newblock \bibinfo{title}{Memristor-the missing circuit element}.
\newblock \bibinfo{journal}{IEEE Transactions on circuit theory} \bibinfo{volume}{18}, \bibinfo{pages}{507--519}.
\newblock {DOI: 10.1109/TCT.1971.1083337}.
%Type = Article

\bibitem[{Lopez-Richard et~al.(2023)Lopez-Richard, Silva, Lipan and Hartmann}]{LopezRichard2022}
\bibinfo{author}{Lopez-Richard, V.}, \bibinfo{author}{Silva, R.S.W.}, \bibinfo{author}{Lipan, O.}, \bibinfo{author}{Hartmann, F.}, \bibinfo{year}{2023}.
\newblock \bibinfo{title}{Tuning the conductance topology in solids}.
\newblock \bibinfo{journal}{Journal of Applied Physics} \bibinfo{volume}{133}, \bibinfo{pages}{134901}.
\newblock {DOI: 10.1063/5.0142721}.
%Type = Article

\bibitem[{Ascoli et~al.(2016)Ascoli, Tetzlaff, Chua, Strachan and Williams}]{ascoli2016history}
\bibinfo{author}{Ascoli, A.}, \bibinfo{author}{Tetzlaff, R.}, \bibinfo{author}{Chua, L.O.}, \bibinfo{author}{Strachan, J.P.}, \bibinfo{author}{Williams, R.S.}, \bibinfo{year}{2016}.
\newblock \bibinfo{title}{History erase effect in a non-volatile memristor}.
\newblock \bibinfo{journal}{IEEE Transactions on Circuits and Systems I: Regular Papers} \bibinfo{volume}{63}, \bibinfo{pages}{389--400}.
\newblock {DOI: 10.1109/TCSI.2016.2525043}.
%Type = Article

\bibitem[{Bisquert(2024b)}]{bisquert2024inductive}
\bibinfo{author}{Bisquert, J.}, \bibinfo{year}{2024}b.
\newblock \bibinfo{title}{Inductive and capacitive hysteresis of current-voltage curves: Unified structural dynamics in solar energy devices, memristors, ionic transistors, and bioelectronics}.
\newblock \bibinfo{journal}{PRX Energy} \bibinfo{volume}{3}, \bibinfo{pages}{011001}.
%Type = Article

\bibitem[{De~Bastiani et~al.(2016)De~Bastiani, Dell'Erba, Gandini, D'Innocenzo, Neutzner, Kandada, Grancini, Binda, Prato, Ball et~al.}]{de2016ion}
\bibinfo{author}{De~Bastiani, M.}, \bibinfo{author}{Dell'Erba, G.}, \bibinfo{author}{Gandini, M.}, \bibinfo{author}{D'Innocenzo, V.}, \bibinfo{author}{Neutzner, S.}, \bibinfo{author}{Kandada, A.R.S.}, \bibinfo{author}{Grancini, G.}, \bibinfo{author}{Binda, M.}, \bibinfo{author}{Prato, M.}, \bibinfo{author}{Ball, J.M.}, et~al., \bibinfo{year}{2016}.
\newblock \bibinfo{title}{Ion migration and the role of preconditioning cycles in the stabilization of the j-v characteristics of inverted hybrid perovskite solar cells.}
\newblock \bibinfo{journal}{Advanced Energy Materials} \bibinfo{volume}{6}.
%Type = Article

\bibitem[{Shao et~al.(2016)Shao, Yang, Lei, Cao, Yu and Liu}]{shao2016pore}
\bibinfo{author}{Shao, J.}, \bibinfo{author}{Yang, S.}, \bibinfo{author}{Lei, L.}, \bibinfo{author}{Cao, Q.}, \bibinfo{author}{Yu, Y.}, \bibinfo{author}{Liu, Y.}, \bibinfo{year}{2016}.
\newblock \bibinfo{title}{Pore size dependent hysteresis elimination in perovskite solar cells based on highly porous tio2 films with widely tunable pores of 15--34 nm}.
\newblock \bibinfo{journal}{Chemistry of Materials} \bibinfo{volume}{28}, \bibinfo{pages}{7134--7144}.
%Type = Article

\bibitem[{Lemos et~al.(2023)Lemos, Rossato, Ramos, Lima, Affon{\c{c}}o, Trofimov, Michel, Fernandes, Naydenov and Graeff}]{lemos2023electron}
\bibinfo{author}{Lemos, H.G.}, \bibinfo{author}{Rossato, J.H.}, \bibinfo{author}{Ramos, R.A.}, \bibinfo{author}{Lima, J.V.}, \bibinfo{author}{Affon{\c{c}}o, L.J.}, \bibinfo{author}{Trofimov, S.}, \bibinfo{author}{Michel, J.J.}, \bibinfo{author}{Fernandes, S.L.}, \bibinfo{author}{Naydenov, B.}, \bibinfo{author}{Graeff, C.F.}, \bibinfo{year}{2023}.
\newblock \bibinfo{title}{Electron transport bilayer with cascade energy alignment based on nb 2 o 5--ti 3 c 2 mxene/tio 2 for efficient perovskite solar cells}.
\newblock \bibinfo{journal}{Journal of Materials Chemistry C} \bibinfo{volume}{11}, \bibinfo{pages}{3571--3580}.
%Type = Article

\bibitem[{Fernandes et~al.(2022)Fernandes, Garcia, J{\'u}nior, Affon{\c{c}}o, Bagnis, Vila{\c{c}}a, Pontes, Da~Silva and Graeff}]{fernandes2022role}
\bibinfo{author}{Fernandes, S.L.}, \bibinfo{author}{Garcia, L.d.O.}, \bibinfo{author}{J{\'u}nior, R.d.A.R.}, \bibinfo{author}{Affon{\c{c}}o, L.J.}, \bibinfo{author}{Bagnis, D.}, \bibinfo{author}{Vila{\c{c}}a, R.}, \bibinfo{author}{Pontes, F.M.}, \bibinfo{author}{Da~Silva, J.H.}, \bibinfo{author}{Graeff, C.F.}, \bibinfo{year}{2022}.
\newblock \bibinfo{title}{The role of nb2o5 deposition process on perovskite solar cells}.
\newblock \bibinfo{journal}{Journal of Renewable and Sustainable Energy} \bibinfo{volume}{14}.
%Type = Article

\bibitem[{Lopez-Richard et~al.(2024a)Lopez-Richard, Meneghetti~Jr, Nogueira, Hartmann and Graeff}]{LopezRichard2024}
\bibinfo{author}{Lopez-Richard, V.}, \bibinfo{author}{Meneghetti~Jr, L.A.}, \bibinfo{author}{Nogueira, G.L.}, \bibinfo{author}{Hartmann, F.}, \bibinfo{author}{Graeff, C.F.}, \bibinfo{year}{2024}a.
\newblock \bibinfo{title}{Unified model for probing solar cell dynamics via cyclic voltammetry and impedance spectroscopy}.
\newblock \bibinfo{journal}{Physical Review B} \bibinfo{volume}{110}, \bibinfo{pages}{115306}.
%Type = Article

\bibitem[{Green(1982)}]{Green1982}
\bibinfo{author}{Green, M.A.}, \bibinfo{year}{1982}.
\newblock \bibinfo{title}{Solar Cells: Operating Principles, Technology, and System Applications}.
\newblock Prentice-Hall Series in Solid State Physical Electronics, \bibinfo{publisher}{Prentice-Hall}, \bibinfo{address}{Englewood Cliffs, NJ}.
%Type = Article

\bibitem[{Silva et~al.(2022)Silva, Hartmann and Lopez-Richard}]{Silva2022}
\bibinfo{author}{Silva, R.S.W.}, \bibinfo{author}{Hartmann, F.}, \bibinfo{author}{Lopez-Richard, V.}, \bibinfo{year}{2022}.
\newblock \bibinfo{title}{The ubiquitous memristive response in solids}.
\newblock \bibinfo{journal}{IEEE Transactions on Electron Devices} \bibinfo{volume}{69}, \bibinfo{pages}{5351--5356}.
\newblock {DOI: 10.1109/TED.2022.3188958}.
%Type = Article

\bibitem[{de~Paiva et~al.(2022)de~Paiva, Wengenroth~Silva, de~Godoy, Bolaños~Vargas, Peres, Soares and Lopez-Richard}]{Paiva2022}
\bibinfo{author}{de~Paiva, A.B.}, \bibinfo{author}{Wengenroth~Silva, R.S.}, \bibinfo{author}{de~Godoy, M.P.F.}, \bibinfo{author}{Bolaños~Vargas, L.M.}, \bibinfo{author}{Peres, M.L.}, \bibinfo{author}{Soares, D.A.W.}, \bibinfo{author}{Lopez-Richard, V.}, \bibinfo{year}{2022}.
\newblock \bibinfo{title}{Temperature, detriment, or advantage for memory emergence: The case of zno}.
\newblock \bibinfo{journal}{The Journal of Chemical Physics} \bibinfo{volume}{157}, \bibinfo{pages}{014704}.
\newblock {DOI: 10.1063/5.0097470}.
%Type = Article

\bibitem[{Clarke et~al.(2023)Clarke, Cowley, Wolf, Cameron, Walker and Richardson}]{clarke2023inverted}
\bibinfo{author}{Clarke, W.}, \bibinfo{author}{Cowley, M.V.}, \bibinfo{author}{Wolf, M.J.}, \bibinfo{author}{Cameron, P.}, \bibinfo{author}{Walker, A.}, \bibinfo{author}{Richardson, G.}, \bibinfo{year}{2023}.
\newblock \bibinfo{title}{Inverted hysteresis as a diagnostic tool for perovskite solar cells: Insights from the drift-diffusion model}.
\newblock \bibinfo{journal}{Journal of Applied Physics} \bibinfo{volume}{133}.
%Type = Article

\bibitem[{Wu et~al.(2018)Wu, Pathak, Chen, Wang, Bahrami, Zhang and Qiao}]{wu2018inverted}
\bibinfo{author}{Wu, F.}, \bibinfo{author}{Pathak, R.}, \bibinfo{author}{Chen, K.}, \bibinfo{author}{Wang, G.}, \bibinfo{author}{Bahrami, B.}, \bibinfo{author}{Zhang, W.H.}, \bibinfo{author}{Qiao, Q.}, \bibinfo{year}{2018}.
\newblock \bibinfo{title}{Inverted current--voltage hysteresis in perovskite solar cells}.
\newblock \bibinfo{journal}{ACS Energy Letters} \bibinfo{volume}{3}, \bibinfo{pages}{2457--2460}.
%Type = Article

\bibitem[{Gonzales et~al.(2022)Gonzales, Guerrero and Bisquert}]{gonzales2022transition}
\bibinfo{author}{Gonzales, C.}, \bibinfo{author}{Guerrero, A.}, \bibinfo{author}{Bisquert, J.}, \bibinfo{year}{2022}.
\newblock \bibinfo{title}{Transition from capacitive to inductive hysteresis: A neuron-style model to correlate i--v curves to impedances of metal halide perovskites}.
\newblock \bibinfo{journal}{The Journal of Physical Chemistry C} \bibinfo{volume}{126}, \bibinfo{pages}{13560--13578}.
%Type = Book

\bibitem[{Bisquert and Guerrero(2022)}]{bisquert2022chemical}
\bibinfo{author}{Bisquert, J.}, \bibinfo{author}{Guerrero, A.}, \bibinfo{year}{2022}.
\newblock \bibinfo{title}{Chemical inductor}.
\newblock \bibinfo{journal}{Journal of the American Chemical Society} \bibinfo{volume}{144}, \bibinfo{pages}{5996--6009}.
\newblock {DOI: 10.1021/jacs.2c00777}.
%Type = Article

\bibitem[{Hern{\'a}ndez-Balaguera and Bisquert(2024)}]{hernandez2024time}
\bibinfo{author}{Hern{\'a}ndez-Balaguera, E.}, \bibinfo{author}{Bisquert, J.}, \bibinfo{year}{2024}.
\newblock \bibinfo{title}{Time transients with inductive loop traces in metal halide perovskites}.
\newblock \bibinfo{journal}{Advanced Functional Materials} \bibinfo{volume}{34}, \bibinfo{pages}{2308678}.
\newblock {DOI: 10.1002/adfm.202308678}.
%Type = Article

\bibitem[{Assun{\c{c}}{\~a}o et~al.(2024)Assun{\c{c}}{\~a}o, Lemos, Rossato, Nogueira, Lima, Fernandes, Nishihora, Fernandes, Louren{\c{c}}o, Bagnis et~al.}]{assunccao2024interface}
\bibinfo{author}{Assun{\c{c}}{\~a}o, J.P.F.}, \bibinfo{author}{Lemos, H.G.}, \bibinfo{author}{Rossato, J.H.}, \bibinfo{author}{Nogueira, G.L.}, \bibinfo{author}{Lima, J.V.}, \bibinfo{author}{Fernandes, S.L.}, \bibinfo{author}{Nishihora, R.K.}, \bibinfo{author}{Fernandes, R.V.}, \bibinfo{author}{Louren{\c{c}}o, S.A.}, \bibinfo{author}{Bagnis, D.}, et~al., \bibinfo{year}{2024}.
\newblock \bibinfo{title}{Interface passivation with ti 3 c 2 t x-mxene doped pmma film for highly efficient and stable inverted perovskite solar cells}.
\newblock \bibinfo{journal}{Journal of Materials Chemistry C} \bibinfo{volume}{12}, \bibinfo{pages}{562--574}.
\newblock {DOI: 10.1039/D3TC03810F}.
%Type = Article

\bibitem[{Calado et~al.(2016)Calado, Telford, Bryant, Li, Nelson, O’regan and Barnes}]{calado2016evidence}
\bibinfo{author}{Calado, P.}, \bibinfo{author}{Telford, A.M.}, \bibinfo{author}{Bryant, D.}, \bibinfo{author}{Li, X.}, \bibinfo{author}{Nelson, J.}, \bibinfo{author}{O’regan, B.C.}, \bibinfo{author}{Barnes, P.R.}, \bibinfo{year}{2016}.
\newblock \bibinfo{title}{Evidence for ion migration in hybrid perovskite solar cells with minimal hysteresis}.
\newblock \bibinfo{journal}{Nature communications} \bibinfo{volume}{7}, \bibinfo{pages}{13831}.
%Type = Article

\bibitem[{Liu et~al.(2019)Liu, Wang, Liu, Yang and Shao}]{liu2019fundamental}
\bibinfo{author}{Liu, P.}, \bibinfo{author}{Wang, W.}, \bibinfo{author}{Liu, S.}, \bibinfo{author}{Yang, H.}, \bibinfo{author}{Shao, Z.}, \bibinfo{year}{2019}.
\newblock \bibinfo{title}{Fundamental understanding of photocurrent hysteresis in perovskite solar cells}.
\newblock \bibinfo{journal}{Advanced Energy Materials} \bibinfo{volume}{9}, \bibinfo{pages}{1803017}.
%Type = Article

\bibitem[{Liu et~al.(2021)Liu, Hu, Dai, Que, Padture and Zhou}]{liu2021correlations}
\bibinfo{author}{Liu, J.}, \bibinfo{author}{Hu, M.}, \bibinfo{author}{Dai, Z.}, \bibinfo{author}{Que, W.}, \bibinfo{author}{Padture, N.P.}, \bibinfo{author}{Zhou, Y.}, \bibinfo{year}{2021}.
\newblock \bibinfo{title}{Correlations between electrochemical ion migration and anomalous device behaviors in perovskite solar cells}.
\newblock \bibinfo{journal}{ACS Energy Letters} \bibinfo{volume}{6}, \bibinfo{pages}{1003--1014}.
%Type = Article

\bibitem[{Tress(2017)}]{tress2017metal}
\bibinfo{author}{Tress, W.}, \bibinfo{year}{2017}.
\newblock \bibinfo{title}{Metal halide perovskites as mixed electronic-ionic conductors: Challenges and opportunities-from hysteresis to memristivity}.
\newblock \bibinfo{journal}{The journal of physical chemistry letters} \bibinfo{volume}{8}, \bibinfo{pages}{3106--3114}.
%Type = Article

\bibitem[{Weber et~al.(2018)Weber, Hermes, Turren-Cruz, Gort, Bergmann, Gilson, Hagfeldt, Graetzel, Tress and Berger}]{weber2018formation}
\bibinfo{author}{Weber, S.A.}, \bibinfo{author}{Hermes, I.M.}, \bibinfo{author}{Turren-Cruz, S.H.}, \bibinfo{author}{Gort, C.}, \bibinfo{author}{Bergmann, V.W.}, \bibinfo{author}{Gilson, L.}, \bibinfo{author}{Hagfeldt, A.}, \bibinfo{author}{Graetzel, M.}, \bibinfo{author}{Tress, W.}, \bibinfo{author}{Berger, R.}, \bibinfo{year}{2018}.
\newblock \bibinfo{title}{How the formation of interfacial charge causes hysteresis in perovskite solar cells}.
\newblock \bibinfo{journal}{Energy \& Environmental Science} \bibinfo{volume}{11}, \bibinfo{pages}{2404--2413}.
%Type = Article

\bibitem[{Sandberg et~al.(2020)Sandberg, Kurpiers, Stolterfoht, Neher, Meredith, Shoaee and Armin}]{sandberg2020question}
\bibinfo{author}{Sandberg, O.J.}, \bibinfo{author}{Kurpiers, J.}, \bibinfo{author}{Stolterfoht, M.}, \bibinfo{author}{Neher, D.}, \bibinfo{author}{Meredith, P.}, \bibinfo{author}{Shoaee, S.}, \bibinfo{author}{Armin, A.}, \bibinfo{year}{2020}.
\newblock \bibinfo{title}{On the question of the need for a built-in potential in perovskite solar cells}.
\newblock \bibinfo{journal}{Advanced Materials Interfaces} \bibinfo{volume}{7}, \bibinfo{pages}{2000041}.
%Type = Article

\bibitem[{Lopez-Richard et~al.(2024b)Lopez-Richard, Pradhan, Wengenroth~Silva, Lipan, Castelano, H{\"o}fling and Hartmann}]{lopez2024beyond}
\bibinfo{author}{Lopez-Richard, V.}, \bibinfo{author}{Pradhan, S.}, \bibinfo{author}{Wengenroth~Silva, R.S.}, \bibinfo{author}{Lipan, O.}, \bibinfo{author}{Castelano, L.K.}, \bibinfo{author}{H{\"o}fling, S.}, \bibinfo{author}{Hartmann, F.}, \bibinfo{year}{2024}b.
\newblock \bibinfo{title}{Beyond equivalent circuit representations in nonlinear systems with inherent memory}.
\newblock \bibinfo{journal}{Journal of Applied Physics} \bibinfo{volume}{136}.
%Type = Article

\bibitem[{Von~Hauff and Klotz(2022)}]{von2022impedance}
\bibinfo{author}{Von~Hauff, E.}, \bibinfo{author}{Klotz, D.}, \bibinfo{year}{2022}.
\newblock \bibinfo{title}{Impedance spectroscopy for perovskite solar cells: characterisation, analysis, and diagnosis}.
\newblock \bibinfo{journal}{Journal of Materials Chemistry C} \bibinfo{volume}{10}, \bibinfo{pages}{742--761}.
%Type = Article

\bibitem[{Hernández-Balaguera and Bisquert(2024)}]{h2024accelerating}
\bibinfo{author}{Hernández-Balaguera, E.}, \bibinfo{author}{Bisquert, J.}, \bibinfo{year}{2024}.
\newblock \bibinfo{title}{Accelerating the assessment of hysteresis in perovskite solar cells}.
\newblock \bibinfo{journal}{ACS Energy Letters} \bibinfo{volume}{9}, \bibinfo{pages}{478--486}.
%Type = Article

\bibitem[{Hernández-Balaguera and Bisquert(2022)}]{hernandez2022negative}
\bibinfo{author}{Hernández-Balaguera, E.}, \bibinfo{author}{Bisquert, J.}, \bibinfo{year}{2022}.
\newblock \bibinfo{title}{Negative transient spikes in halide perovskites}.
\newblock \bibinfo{journal}{ACS Energy Letters} \bibinfo{volume}{7}, \bibinfo{pages}{2602--2610}.
%Type = Article

\bibitem[{O'Kane et~al.(2017)O'Kane, Richardson, Pockett, Niemann, Cave, Sakai, Eperon, Snaith, Foster, Cameron et~al.}]{o2017measurement}
\bibinfo{author}{O'Kane, S.E.}, \bibinfo{author}{Richardson, G.}, \bibinfo{author}{Pockett, A.}, \bibinfo{author}{Niemann, R.G.}, \bibinfo{author}{Cave, J.M.}, \bibinfo{author}{Sakai, N.}, \bibinfo{author}{Eperon, G.E.}, \bibinfo{author}{Snaith, H.J.}, \bibinfo{author}{Foster, J.M.}, \bibinfo{author}{Cameron, P.J.}, et~al., \bibinfo{year}{2017}.
\newblock \bibinfo{title}{Measurement and modelling of dark current decay transients in perovskite solar cells}.
\newblock \bibinfo{journal}{Journal of Materials Chemistry C} \bibinfo{volume}{5}, \bibinfo{pages}{452--462}.
\newblock {DOI: 10.1039/C6TC04964H}.
%Type = Article

\bibitem[{Schalenbach et~al.(2021)Schalenbach, Durmus, Tempel, Kungl and Eichel}]{schalenbach2021double}
\bibinfo{author}{Schalenbach, M.}, \bibinfo{author}{Durmus, Y.E.}, \bibinfo{author}{Tempel, H.}, \bibinfo{author}{Kungl, H.}, \bibinfo{author}{Eichel, R.A.}, \bibinfo{year}{2021}.
\newblock \bibinfo{title}{Double layer capacitances analysed with impedance spectroscopy and cyclic voltammetry: Validity and limits of the constant phase element parameterization}.
\newblock \bibinfo{journal}{Physical Chemistry Chemical Physics} \bibinfo{volume}{23}, \bibinfo{pages}{21097--21105}.
%Type = Article

\bibitem[{Thiesbrummel et~al.(2024)Thiesbrummel, Shah, Gutierrez-Partida, Zu, Pe{\~n}a-Camargo, Zeiske, Diekmann, Ye, Peters, Brinkmann et~al.}]{thiesbrummel2024ion}
\bibinfo{author}{Thiesbrummel, J.}, \bibinfo{author}{Shah, S.}, \bibinfo{author}{Gutierrez-Partida, E.}, \bibinfo{author}{Zu, F.}, \bibinfo{author}{Pe{\~n}a-Camargo, F.}, \bibinfo{author}{Zeiske, S.}, \bibinfo{author}{Diekmann, J.}, \bibinfo{author}{Ye, F.}, \bibinfo{author}{Peters, K.P.}, \bibinfo{author}{Brinkmann, K.O.}, et~al., \bibinfo{year}{2024}.
\newblock \bibinfo{title}{Ion-induced field screening as a dominant factor in perovskite solar cell operational stability}.
\newblock \bibinfo{journal}{Nature Energy} , \bibinfo{pages}{1--13}.
%Type = Article

\bibitem[{Zuo et~al.(2023)Zuo, Li and Chen}]{zuo2023ion}
\bibinfo{author}{Zuo, L.}, \bibinfo{author}{Li, Z.}, \bibinfo{author}{Chen, H.}, \bibinfo{year}{2023}.
\newblock \bibinfo{title}{Ion migration and accumulation in halide perovskite solar cells}.
\newblock \bibinfo{journal}{Chinese Journal of Chemistry} \bibinfo{volume}{41}, \bibinfo{pages}{861--876}.
%Type = Article

\bibitem[{Gollino et~al.(2024)Gollino, Zheng, Mercier and Pauport{\'e}}]{gollino2024unveiling}
\bibinfo{author}{Gollino, L.}, \bibinfo{author}{Zheng, D.}, \bibinfo{author}{Mercier, N.}, \bibinfo{author}{Pauport{\'e}, T.}, \bibinfo{year}{2024}.
\newblock \bibinfo{title}{Unveiling of a puzzling dual ionic migration in lead-and iodide-deficient halide perovskites (d-hps) and its impact on solar cell j--v curve hysteresis}, in: \bibinfo{booktitle}{Exploration}, \bibinfo{organization}{Wiley Online Library}. p. \bibinfo{pages}{20220156}.
\newblock {DOI: 10.1002/EXP.20220156}.
%Type = Article

\bibitem[{Yuan et~al.(2023)Yuan, Lou, Li, Wang, Wang, Ai and Zhang}]{yuan2023targeted}
\bibinfo{author}{Yuan, S.}, \bibinfo{author}{Lou, F.}, \bibinfo{author}{Li, Y.}, \bibinfo{author}{Wang, H.Y.}, \bibinfo{author}{Wang, Y.}, \bibinfo{author}{Ai, X.C.}, \bibinfo{author}{Zhang, J.P.}, \bibinfo{year}{2023}.
\newblock \bibinfo{title}{Targeted suppression of hysteresis effect in perovskite solar cells through the inhibition of cation migration}.
\newblock \bibinfo{journal}{Applied Physics Letters} \bibinfo{volume}{122}.
\newblock {DOI: 10.1063/5.0145249}.
%Type = Article

\bibitem[{Tammireddy et~al.(2024)Tammireddy, Lintangpradipto, Telschow, Futscher, Ehrler, Bakr, Vaynzof and Deibel}]{tammireddy2024hysteresis}
\bibinfo{author}{Tammireddy, S.}, \bibinfo{author}{Lintangpradipto, M.N.}, \bibinfo{author}{Telschow, O.}, \bibinfo{author}{Futscher, M.H.}, \bibinfo{author}{Ehrler, B.}, \bibinfo{author}{Bakr, O.M.}, \bibinfo{author}{Vaynzof, Y.}, \bibinfo{author}{Deibel, C.}, \bibinfo{year}{2024}.
\newblock \bibinfo{title}{Hysteresis and its correlation to ionic defects in perovskite solar cells}.
\newblock \bibinfo{journal}{The Journal of Physical Chemistry Letters} \bibinfo{volume}{15}, \bibinfo{pages}{1363--1372}.
\newblock {DOI: 10.1021/acs.jpclett.3c03146}.
%Type = Article

\end{thebibliography}
\end{document}

\newpage
\section{Supporting results}

Evaluating $V_{oc}$ at different illumination intensities (Fig. S1a) offers insights into recombination kinetics in PSCs. The relationship between Voc and illumination, shown in Fig. S1b, follows the expression:

\begin{equation}
    V_{oc}=\frac{nkT}{q} ln(I),
    \label{Voc}
\end{equation}
where $n$ is the ideality factor. An ideality factor of 1 indicates that second-order (bimolecular) charge-carrier recombination is dominant, whereas a value of $n = 2$ corresponds to first-order (monomolecular) trap-assisted non-radiative recombination. Our study reveals an ideality factor of 1.65.

We employed Mott-Schottky analysis to determine the built-in potential ($V_{bi}$). To ensure accurate measurements of the depletion layer capacitance, the capacitance-voltage (C-V) characteristics were recorded as shown in Fig. S1c in the dark at three distinct frequencies within the dielectric response range: 10, 55 and 100 kHz. We selected 100 kHz to avoid low-frequency contributions that might obscure the measurement\cite{seo2018novel}. The built-in potential ($V_{bi}$) was calculated using the Mott-Schottky relationship\cite{almora2016mott} from the data shown in Fig. S1d:

\begin{equation}
    C_{dl}^{-2}=\frac{2}{q\epsilon\epsilon_0N} (V_{bi}-V),
    \label{MS}
\end{equation}
where $C_{dl}$ is the depletion layer capacitance and $N$ is the charge density. 

\begin{figure}[H]
  \centering 
  \includegraphics[width=16cm]{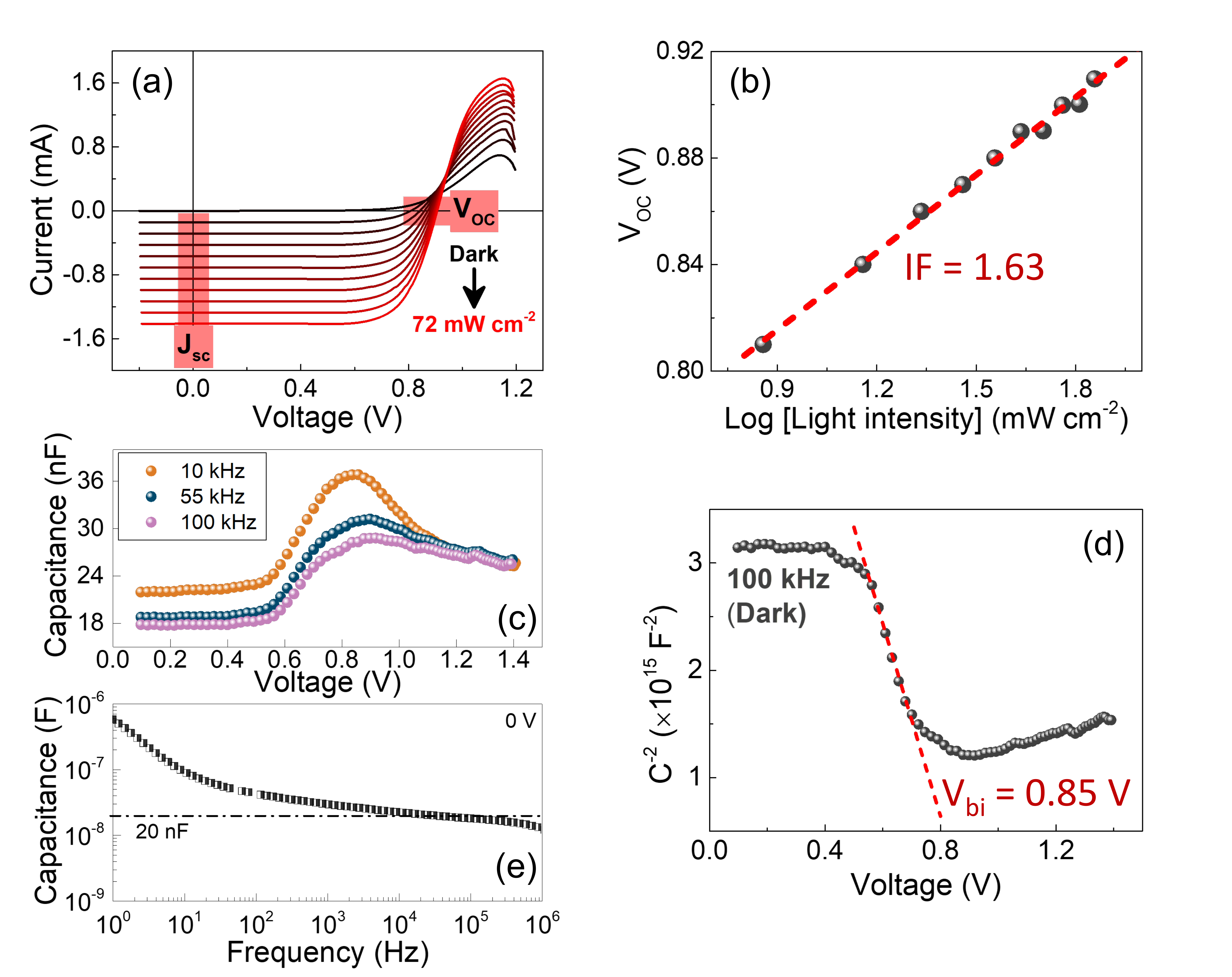}
  \caption{(a) Current-voltage behavior under different illumination intensity, used to calculate the ideality factor. (b) $V_{\text{oc}}$ vs light intensity plot obtained from the current-voltage curves measured from a reverse sweep under different illumination intensities.  (d) Mott-Schottky plot to extract the $V_{\text{bi}}$ from the capacitance vs. voltage with a forward sweep. Plots of capacitance against (c) voltage and (e) frequency. Both measured in the dark.}
  \label{figS1}%
\end{figure}
\clearpage

\begin{figure}
  \centering 
  \includegraphics[width=16cm]{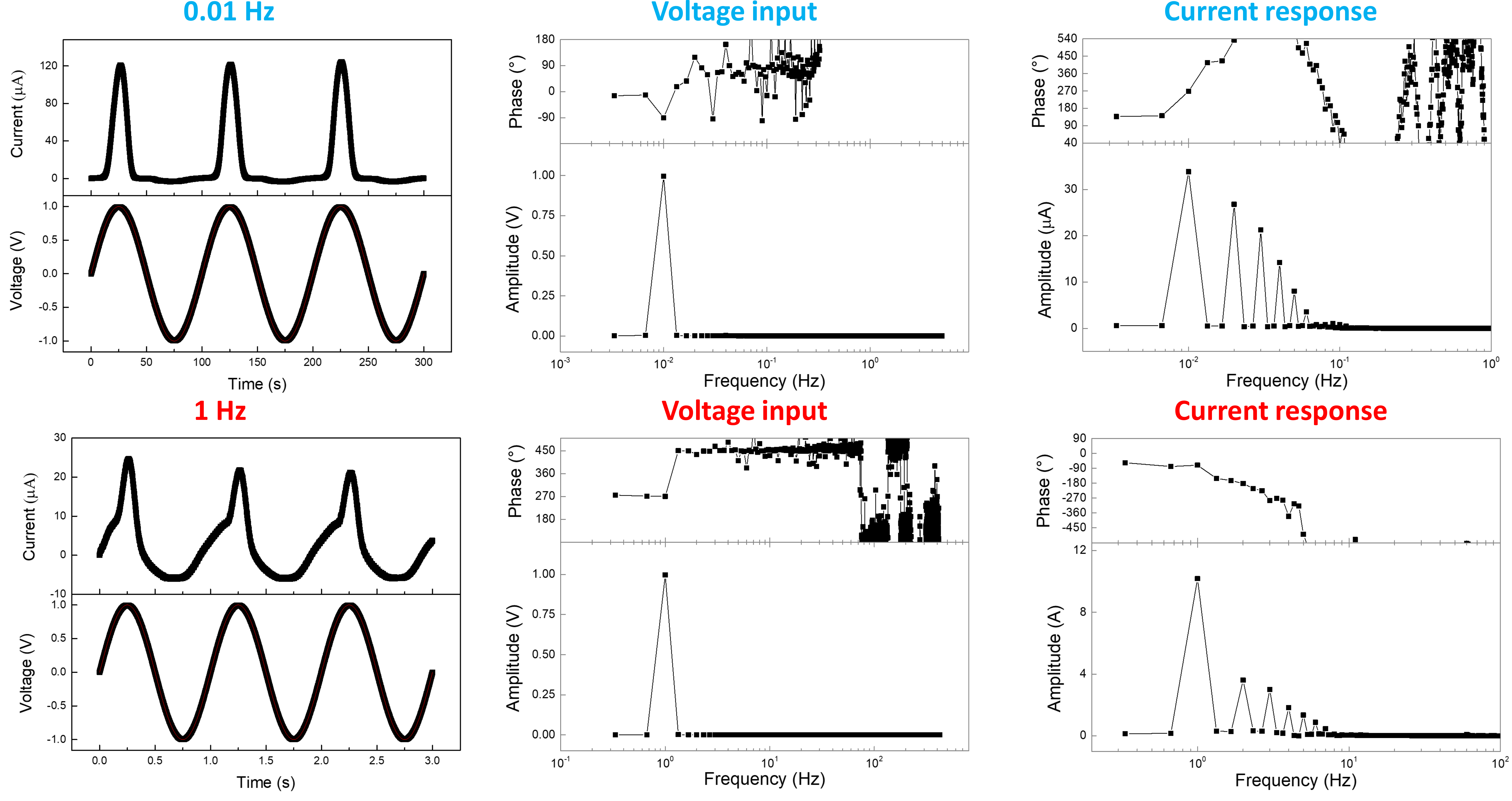}
  \caption{Fast Fourier Transform (FFT) algorithm applied to the experimental sinusoidal data obtained from the n-i-p mesoporous PSC using the PAIOS equipment.}
  \label{figS2}%
\end{figure}

\begin{figure}[H]
  \centering 
  \includegraphics[width=16cm]{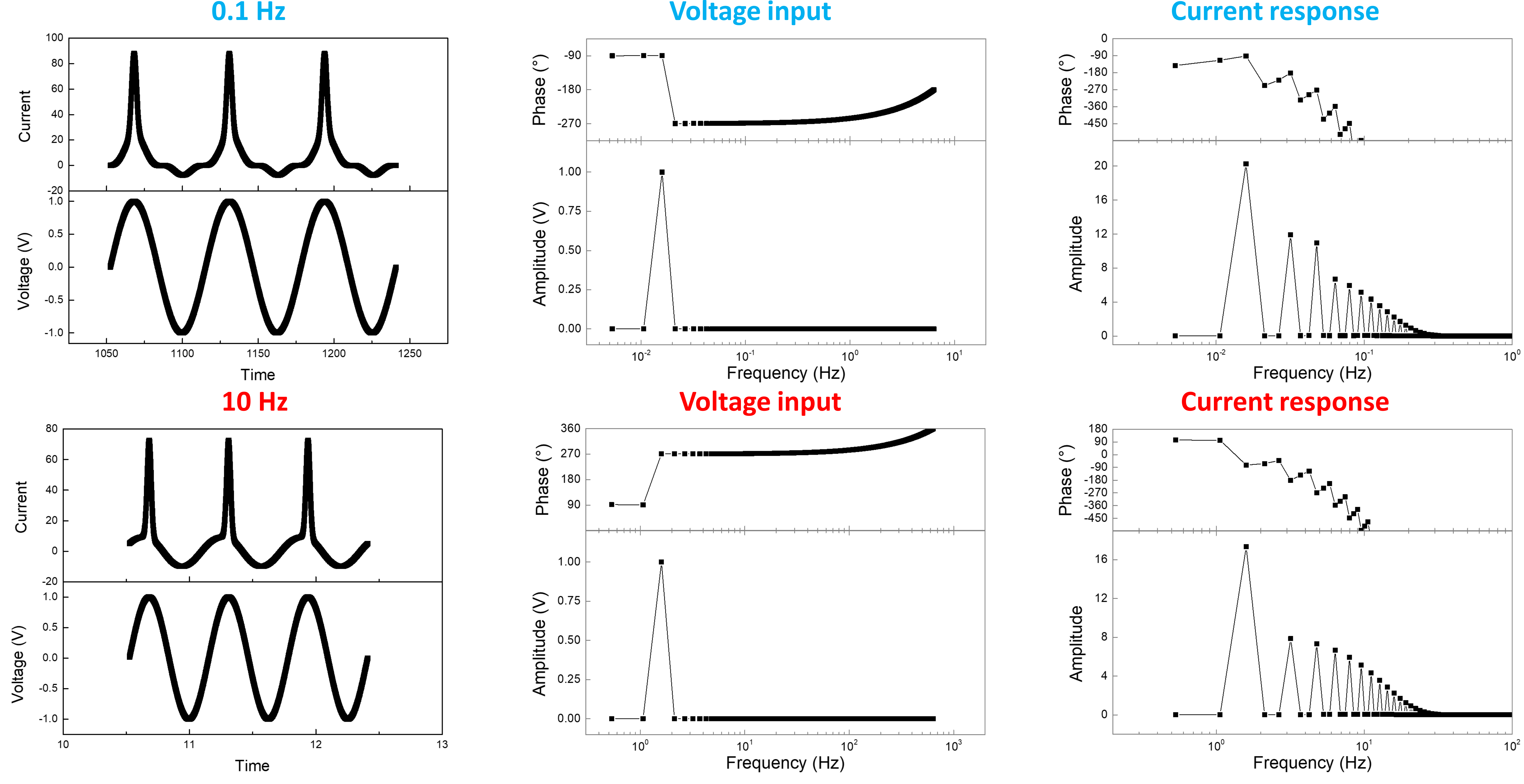}
  \caption{Fast Fourier Transform (FFT) algorithm applied to the simulated sinusoidal data.}
  \label{figS3}%
\end{figure}

\begin{figure}[H]
  \centering 
  \includegraphics[width=16cm]{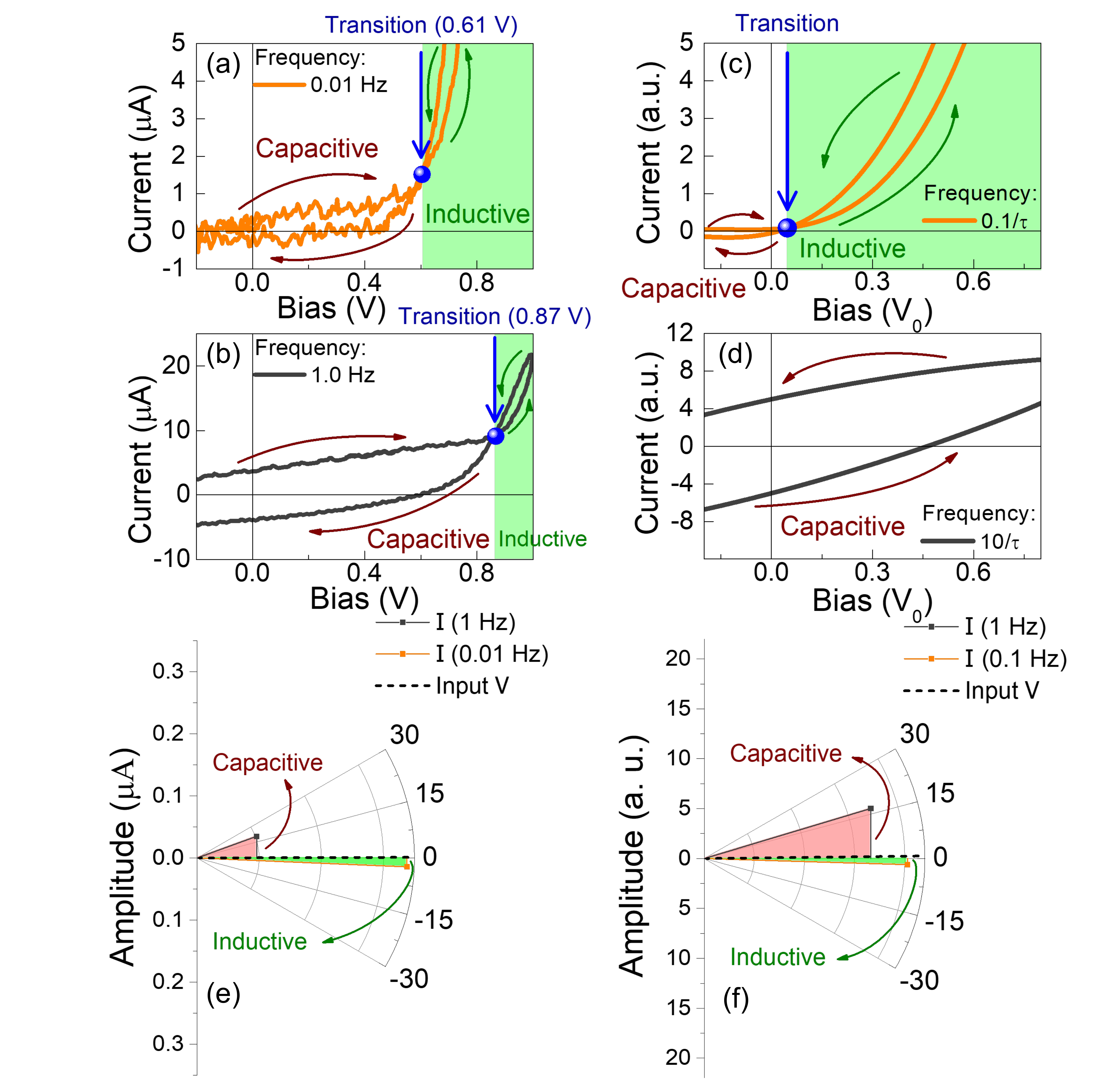}
  \caption{Plots extracted from the Sinus-J-V characterization and simulated data, both for the dark condition. Current-voltage response on the linear scale, with the arrows indicating the scan direction. The experimental data were obtained for (a) 0.01 Hz and (b) 1.0 Hz. The simulation considered the frequencies of (c) $0.1/\tau$ and (d) $10/\tau$. The transition voltage marked by the blue circle separates the capacitive and the inductive-like effects. Polar map highlighting the capacitive and inductive-like effect on the current phase compared to the input voltage, where (e) is the experimental data and (f) the simulated one. }
  \label{figS4}%
\end{figure}

\clearpage

\begin{equation}
    f(x)=A + b_1\exp(-\frac{t}{\tau_1}) + b_2\exp(-\frac{t}{\tau_2}) + b_3\exp(-\frac{t}{\tau_3}),
    \label{exp}
\end{equation}

\begin{table}[H]
    \caption{Parameters used in the exponential function (Eq. S3) to fit the current decay and growth after a voltage pulse.}
    \label{tab:completa}
    \centering
    \begin{tabular}{|c|r|}
        \hline
        Parameters & Values\\
        \hline
        A  & $4.7027\times10{-5}$    \\
        $b_1$  & $7.2800\times10{-5}$    \\
        $b_2$  & $2.9374\times10{-5}$   \\
        $b_3$  & $-1.5893\times10{-5}$   \\
        $\tau_1$  & $3.9287\times10{-5}$    \\
        $\tau_2$  & $4.2046\times10{-4}$   \\
        $\tau_3$  & $1.918\times10{-2}$   \\
        $\chi2$  & $6.2623\times10{-12}$   \\
        \hline
    \end{tabular}
\end{table}